# Conversational Text Extraction with Large Language Models Using Retrieval-Augmented Systems

Soham Roy[1], Mitul Goswami[1], Nisharg Nargund[1], Suneeta Mohanty[1] and Prasant Kumar Pattnaik[1]

*School of Computer Engineering, Kalinga Institute of Industrial Technology, Patia, Bhubaneswar, 751024, India*[1.]

*Abstract*— This study introduces a system leveraging Large Language Models (LLMs) to extract text and enhance user interaction with PDF documents via a conversational interface. Utilizing Retrieval-Augmented Generation (RAG), the system provides informative responses to user inquiries while highlighting relevant passages within the PDF. Upon user upload, the system processes the PDF, employing sentence embeddings to create a document-specific vector store. This vector store enables efficient retrieval of pertinent sections in response to user queries. The LLM then engages in a conversational exchange, using the retrieved information to extract text and generate comprehensive, contextually aware answers. While our approach demonstrates competitive ROUGE values compared to existing state-of-the-art techniques for text extraction and summarization, we acknowledge that further qualitative evaluation is necessary to fully assess its effectiveness in real-world applications. The proposed system gives competitive ROUGE values as compared to existing state-of-the-art techniques for text extraction and summarization, thus offering a valuable tool for researchers, students, and anyone seeking to efficiently extract knowledge and gain insights from documents through an intuitive question-answering interface.

## I. Introduction

The ever-growing volume of digital documents, particularly PDFs, presents a significant challenge: efficiently extracting knowledge from their text-heavy content. Over the years, various tools and techniques have been developed to address this issue, from basic keyword search functionalities to more advanced text mining and natural language processing (NLP) algorithms [1]. Despite these advancements, many solutions still fall short of providing contextually relevant information quickly and accurately. The evolution of artificial intelligence and machine learning, particularly in the form of large language models, has revolutionized this process, enabling more sophisticated and efficient extraction of knowledge from vast repositories of digital documents [2].

Large language models (LLMs) have undergone significant evolution, transforming the landscape of natural language processing (NLP) and information retrieval. While the integration of Retrieval-Augmented Generation (RAG) with LLMs is noteworthy, it is essential to recognize that similar frameworks have been explored in existing literature. This paper aims to build upon these studies by providing a tailored application for document interaction. The advent of machine learning, particularly deep learning, marked a significant leap forward, with models like Word2Vec and GloVe introducing word embeddings that captured semantic relationships between words [3]. Furthermore, Transformers utilize self-attention mechanisms to process and understand text in parallel, rather than sequentially, enabling them to capture long-range dependencies and contextual information more effectively. BERT, with its bidirectional approach, improved the understanding of context within a text, while GPT, with its autoregressive nature, excelled in text generation [4][5]. However, while the use of these models has become widespread, the integration of retrieval augmented generation for targeted PDF interaction remains under-explored. This work focuses on addressing this niche, aiming to bridge this gap by combining large language models with document retrieval in the conversational interface, providing a more tailored application in the domain of document interaction. These advancements in LLMs have significantly enhanced text extraction and data retrieval capabilities. This capability is particularly useful for handling the ever-growing volume of digital documents, enabling efficient knowledge extraction and insight generation [6].

Building on the advancements in LLMs, Retrieval-Augmented Generation (RAG) systems enhance the capability of these models by integrating a retrieval mechanism. RAG combines information retrieval and generative processes to produce highly accurate and contextually relevant responses [7]. In a RAG framework, the system first retrieves relevant passages from a large corpus of documents based on the user's query [8][9]. This combination allows the model to generate responses that are both informed by a broad understanding of language and enriched with precise, relevant details from the retrieved content. This approach significantly improves the model's ability to handle complex queries and extract pertinent information from large datasets, making it an invaluable tool for efficient and accurate knowledge extraction. In this study, the authors introduce a novel approach for text extraction that leverages an LLM system to enhance user interaction with documents via a conversational interface.

## II. Related Work

Recent advancements in document understanding and information extraction have been driven by deep learning techniques. Traditional keyword matching and rule-based methods often struggle with complex documents, while deep learning models provide more robust solutions capable of accurately handling intricate structures. For instance, M. Li et al. introduced the "BiomedRAG" model, which supervises retrieval in the biomedical domain through varied chunk database creation, enhancing prediction accuracy [10]. Similarly, M. D. Cyril Zakka et al. developed the "Almanac" framework, which retrieves medical guidelines and treatment advice, outperforming typical LLMs in factuality, completeness, user preference, and safety [11]. Additionally, K. Yang et al. introduced "LeanDojo," a RAG-based LLM that streamlines theorem proving with comprehensive toolkits and data [12]. P. Lewis et al. explored a fine-tuning recipe for RAG models, leveraging pre-trained parametric and non-parametric memory to improve language development [13].

Z. Feng et al. proposed an iterative retrieval-generation collaborative framework that not only allows for the use of both parametric and non-parametric knowledge but also aids in the discovery of the correct reasoning path via retrieval-generation interactions, which is critical for tasks that require multi-step reasoning [14]. J. Miao et al. demonstrated the development of a specialized ChatGPT model connected with an RAG system that is intended to meet the KDIGO 2023 criteria for chronic kidney disease [15]. In a different domain, H. Li et al. demonstrated the efficacy of leveraging attention processes in neural networks to focus on key areas of material for better question answering in language models [16]. Similarly, Y. Zhang et al. suggested a unique Multi-Modal Knowledge-aware Hierarchical Attention Network (MKHAN) to efficiently leverage a multi-modal knowledge graph (MKG) for explainable medical question answering [17]. However, these approaches are often tailored to specific use cases, lacking the generalizability required for broader document interaction tasks.

Our work builds upon these advancements by presenting a RAG-inspired system for the interactive exploration of user-uploaded PDF documents. We employ advanced sentence embeddings to ensure efficient retrieval of relevant content. By integrating this context into the response generation process of the large language model (LLM), we create a more tailored and contextually rich user experience. This approach allows users to engage in focused conversations that explore the specific content and nuances of the uploaded PDFs, thereby enhancing the effectiveness and relevance of information retrieval and dialogue within the system.

## III. RETRIEVAL AUGMENTED GENERATION

The Retrieval-Augmented Generation (RAG) architecture represents a sophisticated integration of information retrieval (IR) and generative modeling techniques, designed to enhance the precision and relevance of generated responses in natural language processing tasks. The RAG process commences with a robust retrieval component that sifts through a vast corpus of documents to pinpoint relevant passages aligned with the user's query. Traditional IR techniques like TF-IDF and BM25 evaluate the statistical relevance of terms across documents, prioritizing those that closely match the query [18]. Fig. 1 demonstrates the detailed RAG architecture. Advanced methods such as neural retrievers, exemplified by Dense Passage Retrieval (DPR), employ deep learning models to encode documents into dense embeddings, capturing semantic relationships and enhancing contextual understanding.

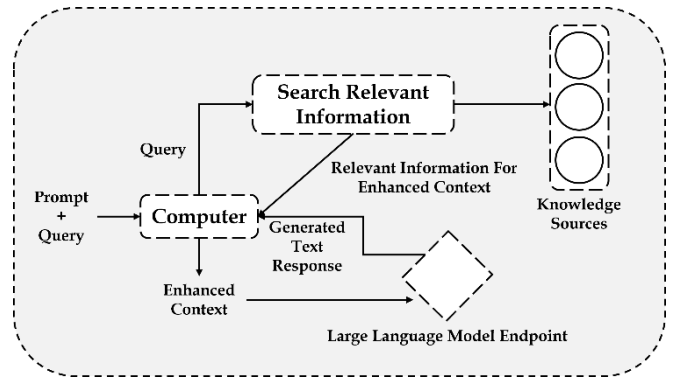

Fig. 1. RAG Architecture

Once relevant passages are identified, they undergo encoding into document embeddings. These embeddings encapsulate the semantic meaning and context of the retrieved text, employing techniques like sentence embeddings from models such as the Universal Sentence Encoder or BERT [19]. These embeddings serve as enriched inputs to the subsequent stage of the RAG architecture. Integration with a generative model, typically an LLM such as GPT, marks the next critical phase. The generative model utilizes the contextual information embedded in the document embeddings to produce responses that are not only grammatically accurate and fluent but also contextually aligned with the user's query [20]. By integrating detailed context from the embeddings, the generative model ensures that its responses are informed by both the broad linguistic knowledge it has learned during training and the specific details extracted from the retrieved passages. To further refine performance, the RAG architecture often involves fine-tuning the generative model on task-specific datasets [21].

## IV. PROPOSED METHODOLOGY

### A. Data Chunking

The integration of the PyPDF2 library enables efficient text extraction and management of PDF documents within the model. Initially, a PdfReader object is created to represent the entire PDF, facilitating seamless interaction with its content [22].

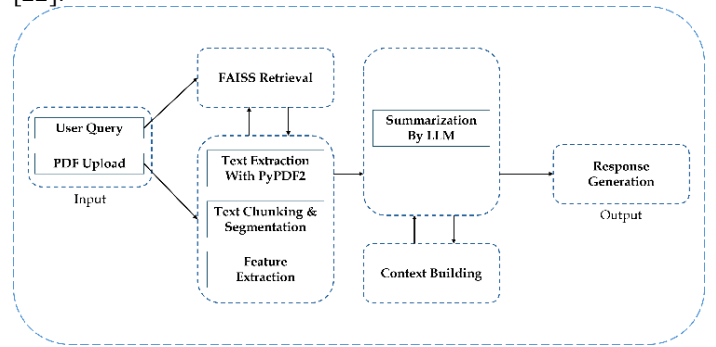

Fig. 2. Workflow of the Proposed Model

To extract the raw text, the model iteratively traverses each page using a loop, employing the extract_text() method. The extracted texts are then consolidated into a single string variable, pdf_text, which captures the entire textual content of the PDF. Given the potentially large size of pdf_text, the model implements text chunking to improve computational efficiency. Equation (1) outlines the mathematical process for text chunking, where n represents the number of desired chunks, chunk_size indicates the size of each chunk, and chunk_overlap defines the overlap between consecutive chunks.

$$i_n = (n - 1) \times (chunk\_size - chunk\_overlap) \quad (1)$$

Equation (1) calculates the starting index for each chunk based on its position n, chunk size, and overlap, ensuring that each chunk overlaps with the previous ones by a specified number of characters. To determine the ending index for each chunk, Equation (2) provides a formula where $j_n$ indicates the ending index of chunk n.

$$j_n = i_n + chunk\_size \quad (2)$$

This approach allows for the systematic division of large text into smaller segments, facilitating easier processing and analysis in natural language processing tasks such as information retrieval, text summarization, and machine translations. Each chunk is associated with metadata to enrich the context and facilitate easier retrieval of specific text segments. Metadata is organized as a list of dictionaries, with each dictionary corresponding to a chunk in the text list. Typically, metadata includes a key-value pair where the key denotes the origin or source identifier of the chunk within the PDF signifies the page number, and "pl" denotes paragraph level). This approach allows for precise tracking and retrieval of information within the PDF document, enhancing the model's capability to handle and manipulate textual content effectively in various applications and user interactions.

*B. Vector Embeddings For Efficient Retrieval*

In preparing text for efficient retrieval, the model utilizes sentence embedding techniques to convert text chunks into numerical representations. This step is crucial for enabling fast and accurate retrieval of semantically similar sentences or passages from a document. Sentence embedding techniques are designed to map sentences from their original high-dimensional textual space into a lower-dimensional vector space. This transformation allows for efficient comparison and retrieval of sentences based on their semantic content. A widely used approach for generating these embeddings is employing pre-trained sentence transformers. In the model, the specific embedding used is "*sentence-transformers/all-MiniLM-L6-v2*". These models are trained on extensive text corpora and have learned to encapsulate the semantic essence of sentences within vector representations [23]. The sentence embedding function is defined in Equation (3) where $S$ is a sentence composed of a sequence of words, $W$ is the embedding matrix for the vocabulary, and $f$ is the sentence embedding function that maps a sentence $S$ to a vector $s \in R^d$.

$$s = f(s) \quad (3)$$

This function $f$ often involves multiple steps, including word embeddings, contextual embeddings using transformer models, and aggregation methods. Each word $\omega_i$ in the sentence $S$ is mapped to a vector $w_i$ using an embedding matrix $W$ in Equation (4) where $W[\omega_i]$

$$w_i = W[\omega_i] \quad (4)$$

To obtain a single fixed-dimensional vector representing the entire sentence, an aggregation method using mean pooling is applied to the contextual embeddings. Equation (5) computes the average of the contextual embeddings of all words in the sentence, resulting in the final sentence embedding

$$S = \frac{1}{n}\sum_1^n h \quad (5)$$

The vector S in Equation (5) is the sentence embedding, which captures the meaning of the sentence in a way that allows for efficient comparison and retrieval in natural language processing tasks. The model uses HuggingFaceEmbedings class from langchain-community.embeddings module to work with the pre-trained sentence transformer model. To load the model, the model name is specified along with any extra configuration options. The embeddings object generates vector representations for each text chunk using the compute_embeddings method, which takes a list of chunks as input and outputs corresponding embedding vectors that capture their meanings in numerical form. These vectors are then combined with metadata, which includes information about the source of each chunk within the PDF document. This integration results in a final list of document representations optimized for efficient retrieval. Consequently, the model can quickly and accurately locate relevant passages in response to user queries, leveraging the meanings captured in the embeddings along with contextual details from the metadata.

*C. Building The Conversational Retrieval Chain(CRC)*

This comprehensive approach involves several key components that synergize to deliver a seamless user experience.

- *Large Language Model*

At the heart of the CRC is the LLM, which generates responses to user queries. The model utilizes the Groq LLM (llm_groq), integrated through the langchain_groq library. This pre-trained LLM leverages its extensive knowledge base, derived from vast amounts of text data, to understand and answer user questions accurately. The LLM's capability to generate coherent and contextually appropriate responses makes it a crucial component of the CRC.

- *Retriever*

The retriever component is responsible for fetching relevant information from the document based on the user's query. The model employs the faiss library from langchain_community.vectorstores to create a vector store using the document embeddings generated earlier. These embeddings transform text chunks into numerical representations that capture their semantic content. The vector store allows for efficient retrieval of document sections (chunks) that are semantically similar to the user's query. The as_retriever method of the vector store object is used to create a retriever object that integrates into the CRC, enabling precise and relevant information retrieval.

- *Memory*

Memory management is essential for preserving conversational context. The model employs the ConversationBufferMemory class from langchain.memory to store past user queries and LLM responses. This history is crucial for the LLM to reference prior interactions when generating current responses. The memory is set up with keys: memory_key="chat_history" for conversation history and output_key="answer" for the LLM's responses. This configuration facilitates more coherent and contextually aware interactions.

- *Chain Configuration*

To integrate these components, the ConversationalRetrievalChain.from_llm method is employed with specific parameters. The LLM parameter is configured to utilize thr Groq LLM object(llm_groq). The chain_type is designated as "stuff", indicating a focus on retrieving factual content from the document. The retriever parameter is linked to the retriever object generated from the vector store, ensuring efficient retrieval of relevant document section. The memory parameter is set to conversation buffer memory object. Further, return_source_documents is True instructing chains to return chunks along with responses. This ensures the accurate answers enriched with relevant context from the documents.

*D. User Interaction And Model Response*

The model enables a natural and interactive conversation between users and their uploaded PDF documents. The process starts when users input their questions through a text field integrated into the Streamlit interface, ensuring that initiating queries is straightforward and accessible. Users type their questions into the provided input field (st.text_input), and upon submission, the system promptly captures the query and processes it using the retrieval chain. The chain.invoke method efficiently directs the query to subsequent stages of the workflow.

At the core of the model's functionality is the Conversational Retrieval Chain. To generate contextually rich responses, the chain first accesses the conversation history via the ConversationBufferMemory [25], which retains past user queries and responses, ensuring that the current interaction benefits from previous exchanges.

Subsequently, the system utilizes a retriever that operates on a pre-constructed document vector store, comprising embeddings of text chunks extracted from the PDF. The retriever searches for document sections that are semantically similar to the user's query, using cosine similarity to evaluate how closely related two vectors are within the embedding space. Equation (6) illustrates the concept of cosine similarity.

$$\cos(u, v) = \frac{u.v}{(\|u\|\|v\|)} \quad (6)$$

Cosine similarity scores range from -1 (completely dissimilar) to 1 (identical). The model retrieves the document sections with the highest cosine similarity scores, indicating their relevance to the user's query. These retrieved sections, along with the conversation context, are then fed into the Groq LLM. By leveraging its pre-trained knowledge and the specific context from the retrieved text, the Groq LLM generates comprehensive and accurate responses to user questions.

When relevant document sections are retrieved, the model enhances responses by referencing these sources. It assigns unique identifiers to each retrieved section and may include these references in the response text. This method not only ensures transparency but also allows users to verify the information's origin. To improve usability, Streamlit expanders (st.expander) are used, enabling users to easily view the content of the retrieved document sections. By clicking on the corresponding source names, users can expand and read the specific excerpts that informed the LLM's response. This interactive feature allows users to explore the document content more deeply, enhancing their understanding and engagement with the material.

V. EXPERIMENTATIONS AND RESULTS

To assess the model's capability to navigate and summarize complex academic materials, we employed ROUGE (Recall-Oriented Understudy for Gisting Evaluation) scores, a widely accepted metric for evaluating the quality of automatically generated summaries against human-written references. However, relying solely on ROUGE metrics may not adequately reflect the system's interactive and conversational aspects. Therefore, future studies will include qualitative evaluations to examine user interaction quality and the model's effectiveness from an end-user perspective, ensuring a comprehensive assessment of its performance.

The evaluation utilized a carefully curated dataset comprising top-cited research articles, known for their dense information content, technical language, and intricate methodologies. These articles posed significant challenges, making them well-suited for rigorously testing the model's summarization capabilities. The article abstracts served as input reference summaries for calculating the ROUGE scores for each document. This analysis provided valuable insights into the model's proficiency in accurately capturing and summarizing critical findings from highly technical and detailed research literature. ROUGE measures the overlap of n-grams between the generated text and the reference text, mathematically represented in Equation (6), where the maximum number of n-grams co-occurring in both the candidate and reference summaries is considered.

$$ROUGE = \frac{\sum_{S \in \{Summaries\}} \sum_{gram_n \in S} Count\_match(gram_n)}{\sum_{S \in \{Summaries\}} \sum_{gram_n \in S} Count(gram_n)} \quad (6)$$

The study specifically used ROUGE-1 (unigrams) and ROUGE-2 (bigrams) in the evaluation. ROUGE-L measures the longest common subsequence (LCS) between the candidate and reference summaries. It's calculated using Equations (7), and (8) followed by Equation (9), where $X$ is the reference summary of length $m$, $Y$ is the candidate summary of length *n*, and *β* is typically set to favor recall (*β* >1).

$$ROUGE - L_{Recall} = \frac{LCS(X,Y)}{m} \quad (7)$$

$$ROUGE - L_{Precision} = \frac{LCS(X,Y)}{n} \quad (8)$$

$$ROUGE - L_F = \frac{(1+\beta^2) \cdot R \cdot P}{R + \beta^2 \cdot P} \quad (9)$$

To evaluate the model performance, the authors tested it with a custom dataset of various research papers sourced from top research databases and analyzed the ROUGE scores of the generated answers. The relatively moderate ROUGE scores can be attributed to the model's focus on condensing extensive content into concise responses. This indicates the model's tendency to prioritize brevity and specificity over word-for-word overlap. The average representative scores obtained from evaluating upon the dataset, are given in Table. I.

TABLE I. PERFORMANCE METRICS OF THE MODEL

| Performance Metric | Score Values (Average) |
|---|---|
| ROUGE – 1 | 0.4604 |
| ROUGE – 2 | 0.3576 |
| ROUGE - L | 0.4283 |

The scores indicate that approximately 46% of individual words (ROUGE-1) and around 35% of bigram phrases (ROUGE-2) in the generated responses matched those found in the original documents. The ROUGE-L score, which lies between ROUGE-1 and ROUGE-2, demonstrates some preservation of word order while accommodating gaps and rephrasing. The relatively low ROUGE scores highlight the system's capability to distill information into concise answers instead of replicating large text segments. Moreover, the complexity and dense information structure of the input research articles creates a high bar for any model aiming to balance conciseness with informativeness. Good summaries often rephrase ideas, leading to lower word-for-word matches but potentially better conveyance of key concepts. Furthermore, the system focuses on providing specific answers to questions, naturally leading to lower overlap with the full text of the documents. Moreover, the significant length difference between focused answers and entire articles further contributes to the lower ROUGE scores. Table. II compares the model performance with other SOTA approaches.

TABLE II. COMPARISON OF PROPOSED MODEL PERFORMANCE METRICS

| Model | ROUGE - 1 | ROUGE - 2 | ROUGE - L |
|---|---|---|---|
| **RAG-PDF (Our Model)** | 0.4604 | 0.3576 | 0.4283 |
| ML + RL ROUGE + Novel, With LM [26] | 0.4019 | 0.1738 | 0.3752 |
| COSUM [27] | 0.4908 | 0.2379 | 0.2834 |
| Latent Semantic Analysis [28] | 0.4621 | 0.2618 | 0.3479 |
| EdgeSumm [29] | 0.5379 | 0.2858 | 0.4979 |
| Generative Adversarial Network [30] | 0.3992 | 0.1765 | 0.3671 |
| TFRSP [31] | 0.2483 | 0.2874 | 0.2043 |

Table II presents a comparative analysis of various models based on ROUGE-1, ROUGE-2, and ROUGE-L scores, which evaluate summary quality against reference summaries. The RAG-PDF model demonstrates strong performance, achieving a ROUGE-1 score of 0.4604, ROUGE-2 score of 0.3576, and ROUGE-L score of 0.4283, indicating its effectiveness in capturing both individual words and longer sequences for coherent summaries.

While EdgeSumm excels in ROUGE-1 and ROUGE-L, its lower ROUGE-2 score reveals limitations in bigram coherence. Our model balances coherence, particularly in complex technical text. In contrast, the ML + RL ROUGE + Novel model shows poorer performance, especially in ROUGE-2 (0.1738) and ROUGE-L (0.3752), suggesting challenges in capturing bi-gram sequences. COSUM performs well in ROUGE-1 (0.4908) but lacks coherence in longer sequences with lower ROUGE-2 (0.2379) and ROUGE-L (0.2834).

Latent Semantic Analysis is comparable to our model in ROUGE-1 (0.4621) but falls short in ROUGE-2 (0.2618) and ROUGE-L (0.3479). The Generative Adversarial Network model exhibits low scores across metrics, particularly in ROUGE-2 (0.1765). Lastly, the TFRSP model scores the lowest in ROUGE-1 (0.2483) and ROUGE-L (0.2043), indicating significant challenges in summary generation.

While ROUGE metrics provide useful insights, they may not fully capture user experience or interaction quality. Therefore, future work will focus on incorporating user-centered evaluations, including qualitative feedback and interaction analysis, to align the system's performance with real-world needs.

VI. CONCLUSION AND FUTURE WORK

The model introduces a unique approach for interacting with PDF documents via a conversational interface, harnessing the power of LLMs and RAG. This system enables users to extract valuable insights from complex and text-heavy materials effectively. One of its standout features is its focus on the specific content of uploaded PDFs, rather than relying on extensive external knowledge bases. By employing sentence embeddings, the model converts text chunks into numerical vectors and utilizes cosine similarity for efficient retrieval, aligning responses closely with user intent. Performance evaluations reveal competitive ROUGE scores—0.4604 for ROUGE-1, 0.3576 for ROUGE-2, and 0.4283 for ROUGE-L—demonstrating the model's capability to capture essential content and structure while outperforming many existing models in summarization and question answering.

To enhance the practical application of this system, future work will aim to generalize its approach for a wider variety of document types. This will include refining the retrieval mechanism to accommodate diverse structures, such as legal, financial, and multimodal documents, thereby increasing the system's versatility in real-world scenarios. We also plan to incorporate reinforcement learning techniques to improve user interactions, allowing the model to adapt dynamically based on feedback. Exploring the incorporation of knowledge graphs and ontologies may also improve semantic understanding and contextualization. Furthermore, refining the model with user interaction data and reinforcement learning can facilitate more personalized responses, ensuring that the system continuously evolves to meet user needs effectively.